\documentstyle[12pt, epsf]{article}
 \textwidth
 16.4cm
 \oddsidemargin
 2.5cm
 \advance\oddsidemargin by
 -1in
 \evensidemargin
 0.0cm
 \advance\evensidemargin
 by
 -1in
 \marginparwidth
 1.9cm
 \marginparsep
 0.4cm
 \marginparpush
 0.4cm
 \topmargin
 -0.5cm
 \advance\topmargin
 by
 -0.0in
 \textheight
 23.0cm
 \makeindex

 \newcommand\beq{\begin{equation}}
 \newcommand\eeq{\end{equation}}
 \newcommand\beqn{\begin{eqnarray}}
 \newcommand\eeqn{\end{eqnarray}}
 \newcommand{\la}{\langle}
 \newcommand{\ra}{\rangle}

\begin{document}

\vspace*{3cm}

\centerline{\Large\bf The Coulomb Phase Revisited}

\vspace{.5cm}

\begin{center} 
{\large B.Z.~Kopeliovich$^{1,2}$ and
A.V.~Tarasov$^{3,1,2}$} \\[8pt]
 $^1${\sl Max-Planck Institut f\"ur Kernphysik, Postfach 103980, 69029
Heidelberg, Germany}\\

$^2${\sl Joint Institute for Nuclear Research, Dubna, 141980 Moscow Region,
Russia}

$^3${\sl Inst. f\"ur Theor. Physik der Universit\"at,    
Philosophenweg 19, 69120 Heidelberg, Germany}
\end{center}

\vspace{2cm}
\begin{abstract}

Motivated by the forthcoming data from the E950 experiment at BNL for
small angle polarized proton-carbon scattering we revisit the problem of
Coulomb phase. The approximation usually used, with the momentum transfer
squared ($q^2$) small compared to the inverse elastic slope, is not
justified within the kinematics of the E950 experiment. We go beyond this
approximation and derive a new rather simple expression which recovers the
formula of Bethe at small $q^2$, but is valid at any momentum transfer.

\end{abstract}

\newpage


The forthcoming precise data for small angle polarized proton-carbon
scattering measured in the E950 experiment at BNL raise again the problem
of Coulomb phase \cite{bethe}. The experiment covers the range of momentum
transfer up to $q^2=0.05\,GeV^2$ where the product $Bq^2=3$ ($B\approx
60\,GeV^2$ is the elastic slope). At the same time the early calculations
of the Coulomb phase neglected terms of the order of $Bq^2$
\cite{bethe,wy}. The next-to-leading corrections of the order of
$q^2\ln(q^2)$ were calculated by Cahn \cite{cahn}. However,
next-to-next-to-leading order terms might be also important at higher
$q^2$.

The asymmetry of polarized proton-nucleus scattering arising from
Coulomb-nuclear interference \cite{cni}, the main goal of experiment E950,
can be essentially affected by the Coulomb phase, especially if the
hadronic spin-flip amplitude has a real part \cite{longpaper}. This
situation motivates us to attack once again the problem of the Coulomb
phase to reveal the corrections of the order of $q^2$ and higher. This has
recently been done numerically for $pp$ scattering in \cite{selyugin}.

We demonstrate that one can arrive at a rather simple analytical
expression Eq.~(\ref{330}) for the Coulomb phase which is valid at any
$q^2$, provided that the electromagnetic formfactor and nuclear amplitude
have Gaussian dependence on the momentum transfer $q$. This assumption is
well justified for light and medium heavy nuclei. For a proton
target the Gaussian formfactor falls off too steeply at $q^2 >
\Lambda^2$, where $\Lambda^2=0.71\,GeV^2$ is the parameter of the 
dipole parameterization. However, the Coulomb phase matters only in the
Coulomb-nuclear interference region $q^2 \ll 0.1\,GeV^2$ where the Gaussian
parameterization is precise.

If the two slopes of $q^2$ dependence for electromagnetic and hadronic 
amplitude coincide the expression for the Coulomb
phase becomes especially simple as is given by Eq.~(\ref{340}).

\begin{figure}[tbh] 
\includegraphics{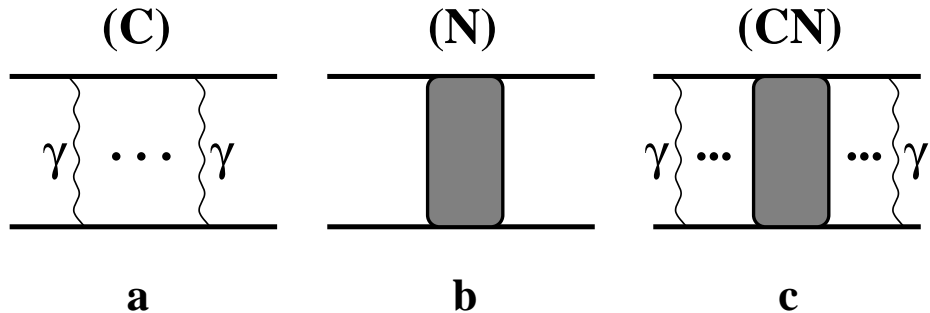} 
\begin{center} 
\vspace{4cm} 
\parbox{13cm}
{\caption[Delta]
{\sl Three types of interaction: pure Coulomb (a),
nuclear (b) and nuclear-Coulomb (c).}
\label{graphs}} 
\end{center} 
\end{figure}

The elastic scattering amplitude can be represented as,
 \beq
f(q)=f_C(q) + f_{NC}(q)\ ,
\label{10}
\eeq
 where the first and second terms correspond to the first (a) and  sum of
the second and third (b,c) graphs depicted in Fig.~\ref{graphs},
respectively. 
Of course such a classification is conventional and we make it for
convenience.

The amplitudes in Eq.~(\ref{10}) read,
\beq
f_C(q)=\frac{i}{2\pi}\int d^2b\,e^{i\vec q\cdot\vec b}\,
\left(1 - e^{i\chi_C(b)}\right)\ ,
\label{20}
\eeq
\beq
f_{NC}(q) = \frac{i}{2\pi}\int d^2b\,
e^{i\vec q\cdot\vec b}\, 
e^{i\chi_C(b)}\, \gamma_N(b)\ ,
\label{30}
\eeq
where
\beq
\gamma_N(b)=\frac{i}{2\pi}\int d^2q\,e^{-i\vec q\cdot\vec b}\,
f_N(q)\ .
\label{40}
\eeq

The amplitudes Eqs.~(\ref{10}) - (\ref{40}) are normalized according
to the conditions,
 \beqn
\frac{d\sigma}{dq^2} &=& \Bigl|f(q)\Bigr|^2\nonumber\\
\sigma_{tot} &=& 4\pi\,{\rm Im}\,f_N(q=0)\ .
\label{50}
\eeqn

In Eqs.~(\ref{10}) - (\ref{50}) $f_C(q)$ and $f_N(q)$ are the net Coulomb
(long-range) and nuclear (proton-nucleus) amplitudes (Figs.~\ref{graphs}a and
b respectively), and $f_{NC}(q)$ includes the effects of strong interaction
(Fig.~\ref{graphs}b) and Coulomb-nuclear interference (Fig.~\ref{graphs}c).

The Coulomb phase $\chi_C(b)$ in Eqs.~(\ref{10})-(\ref{40}) is a Fourier
transform of the Coulomb part of the amplitude calculated in the Born
approximation,
 \beqn
\chi_C(b) &=& \frac{1}{2\pi}\int d^2q\,
e^{-i\vec q\cdot\vec b}\,
f_C^{(B)}(q)\ ;
\label{60}\\
f_C^{(B)}(q) &=& 
-\frac{2\,\alpha Z_1\,Z_2}{q^2+\lambda^2}\,
S(q^2)\ ;
\label{70}\\
S(q^2) &=& F^{(1)}_{em}(q^2)\,F^{(2)}_{em}(q^2)\ ,
\label{80}
\eeqn
 where $F^{(1,2)}_{em}(q^2)$ are the electromagnetic formfactors of the
colliding particles (nuclei), and $Z_{1,2}$ are their charges.
In order to keep the integrals finite we give to the photon a small mass
$\lambda$ which will disappear from the final expressions in the limit
$\lambda\to 0$.

On the contrary to the Born approximation, the full Coulomb amplitude
including
all the multi-photon exchanges is complex, {\it i.e.} has a phase
$\Phi_C(q)$,
 \beq 
f_C(q)=-{\rm sign}(Z_1\,Z_2)\,
\Bigl|f_C(q)\Bigr|\,e^{i\,\Phi_C(q)}\ ,
\label{90}
\eeq
where
 \beq
\Bigl|f_C(q)\Bigr| = \Bigl|f_C^{(B)}(q)\Bigr|\,
\left[1 + O\Bigl((\alpha Z_1Z_2)^2\Bigr)\right]\ .
\label{100}
\eeq

For the sake of simplicity we restrict our consideration to the case of
$\alpha |Z_1Z_2|\ll 1$ (appropriate for $pC$) and will neglect the higher
order
corrections. In this approximation, 
 \beqn
\Phi_C(q) &=& {1\over2}\,\frac{ 
\int d^2b\,\Bigl|\chi_C(b)\Bigr|^2\, 
e^{i\vec q\cdot\vec b}} 
{\int d^2b\,\chi_C(b)\,
e^{i\vec q\cdot\vec b}}\nonumber\\
&=& \frac{1}{2\pi\,f_C^{(B)}(q)}\,
\int d^2q_1\,d^2q_2\,f_C^{(B)}(q_1)\,
f_C^{(B)}(q_2)\,\delta(\vec q-\vec q_1-\vec q_2)\ .
\label{110}
\eeqn

With the same accuracy the amplitude $f_{NC}(q)$ in Eq.~(\ref{30})
can be represented in the form,
 \beq
f_{NC}(q) = f_N(q)\,e^{i\Phi_{NC}(q)} + 
O\Bigl((\alpha Z_1Z_2)^2\Bigr)\ ,
\label{120}
\eeq
where
\beq
f_{NC}(q)\,\Phi_{NC}(q) = 
\frac{1}{2\pi}\,\int d^2q_1\,d^2q_2\,
\delta(\vec q-\vec q_1-\vec q_2)\ .
\label{130}
\eeq

Our results for $\Phi_C$, $\Phi_{NC}$ and $\Delta\Phi=\Phi_C-\Phi_{NC}$
look rather simple if the $q$-dependence of the formfactors in
Eq.~(\ref{30})  is Gaussian,
 \beqn
S(q^2) &=& e^{-a\,q^2}\ ,\ \ \ \ 
a = {1\over6}\,(\la r^2\ra_1 + \la r^2\ra_2)\ ,
\label{140}\\
f_N(q) &=& f_N(0)\,e^{-b\,q^2}\ ,\ \ \ \ \ b=B/2\ ,
\label{150}
\eeqn
where $B$ is the slope of the hadronic differential cross section.

In this case the phase Eq.~(\ref{130}) takes the form,
 \beq
\Phi_{NC}(q) = - \frac{\alpha Z_1Z_2}{\pi}
\int \frac{d^2q_1}{q_1^2+\lambda^2}\,
\exp\Bigl[-(a+b)\,q_1^2 + 2\,b\,\vec q_1\cdot\vec q\Bigr]\ .
\label{160}
\eeq
This integration can be performed analytically replacing
 \beq
\frac{1}{q_1^2+\lambda^2} = \int\limits_0^\infty dt\,
\exp\Bigl[-t\,(q_1^2+\lambda^2)\Bigr]\ .
\label{170}
\eeq
 Then the phase Eq.~(\ref{160}) takes the form,
 \beq
\Phi_{NC}(q) = - \frac{\alpha Z_1Z_2}{\pi}\,
\int\limits_1^\infty {du\over u}\,
\exp\left[{z\over u} - v\,u + v\right]\ ,
\label{180}
\eeq
 where
\beqn
z&=& \frac{b^2\,q^2}{a+b}\ ,\nonumber\\
v&=& \lambda^2\,(a+b)\ .
\label{190}
\eeqn
 Further, expanding the exponential we get,
\beq
\Phi_{NC}(q) = - \frac{\alpha Z_1Z_2}{\pi}\,
\sum\limits_{k=0}^\infty \frac{z^k}{k!}\,
E_{k+1}(v)\,e^v\ ,
\label{200}
\eeq
 where
 \beqn
E_{k+1}(v) &=& \int\limits_1^\infty 
\frac{du}{u^{k+1}}\,e^{-kv}\ ,\nonumber\\
E_{k+1}(v)|_{v\to 0} &=& {1\over k}\ ,\ \ \ \ 
k > 0\ ,\nonumber\\
{\rm E_1}(v)|_{v\to 0} &=& -\gamma - \ln(v)\ .
\label{210}\eeqn
 Here $\gamma=0.5772$ is the Euler
constant.

Thus at $v<<1$ we arrive at,
 \beq
\Phi_{NC}(q) = \alpha Z_1Z_2\, 
\Bigl[2\,\gamma + \ln(vz) - {\rm Ei}(z)\Bigr] + O(v)\ ,
\label{220}
\eeq
 where ${\rm Ei}(z)=-{\rm E_1}(-z)$ is the integral exponential function.

Coming back to the phase (\ref{110}) we represent the integral in the
numerator in the {\it r.h.s.} as,
 \beqn
I_c(q)&=&\frac{1}{4\pi} 
\int d^2q_1\,d^2q_2\,f_C^{(B)}(q_1)\,
f_C^{(B)}(q_2)\,\delta(\vec q-\vec q_1-\vec q_2)
\nonumber\\ 
&=& \frac{1}{16\pi} \int d^2\Delta\ 
f_C^{(B)}\left(\frac{\vec q+\vec\Delta}{2}\right)\,
f_C^{(B)}\left(\frac{\vec q-\vec\Delta}{2}\right)\ .
\label{230}
\eeqn

Since the product
 \beq
S\left(\frac{\vec q+\vec\Delta}{2}\right)\,
S\left(\frac{\vec q-\vec\Delta}{2}\right) =
\exp\left[-{a\over2}\,(q^2+\Delta^2)\right]
\label{240}
\eeq
 is independent of the angle $\phi$ between the vectors $\vec q$ and $\Delta$
($d^2\Delta = d\Delta^2\,d\phi/2$), one can perform integration over
$\phi$ in (\ref{230}),
 \beqn
I_c(q) &=& \frac{2\,(\alpha Z_1Z_2)^2}{\pi}\,
\int\limits_0^\infty d\Delta^2\,\int\limits_0^{2\pi} d\phi\,
\exp\left[-{a\over2}\,(q^2+\Delta^2)\right]\nonumber\\
&\times&
\Bigl[(q^2+2q\Delta\,\cos\phi +\Delta^2+4\lambda^2)\,
(q^2-2q\Delta\,\cos\phi +\Delta^2+4\lambda^2)\Bigr]^{-1}\ .
\label{250}
\eeqn
Then, it can be represented as,
 \beq
I_c(q) = 2\,(\alpha Z_1Z_2)^2\,
\int\limits_0^\infty d\Delta^2\,
\exp\left[-{a\over2}\,(q^2+\Delta^2)\right]\ 
\frac{\partial\Psi(\Delta^2,q^2,\lambda^2)}
{\partial\Delta^2}\ ,
\label{260}
\eeq
 where
\beq
\Psi(\Delta^2,q^2,\lambda^2) =
\frac{1}{\sqrt{q^2(q^2+\lambda^2)}}\ 
\ln\left[\sqrt{\left(u-{1\over2}\right)^2+
\frac{\lambda^2}{q^2}}
+u-{1\over2}\right]\ ,
\label{270}\\
\eeq
 \beq
u = \frac{q^2+4\lambda^2}{q^2+\Delta^2+4\lambda^2}\ .
\nonumber
\eeq

In the limit $\lambda \to 0$ function $\Psi(\Delta^2,q^2,\lambda^2)$
can be expanded dependent on the relation between $\Delta^2$ and $q^2$,
 \beqn
\Psi(\Delta^2,q^2,\lambda^2)|_{\Delta^2 < q^2} &=& 
{1\over q^2}\,\ln\left(\frac{q^2-\Delta^2}
{q^2+\Delta^2}\right) +O\left({\lambda^2\over q^2}\right)\ ,
\nonumber\\
\Psi(\Delta^2,q^2,\lambda^2)|_{\Delta^2 > q^2} &=& 
{1\over q^2}\,\left[\ln\left(\frac{\lambda^2}{q^2}\right) - 
\ln\left(\frac{\Delta^2-q^2}
{\Delta^2+q^2}\right)\right] + 
O\left({\lambda^2\over q^2}\right)\ ,
\label{280}
\eeqn

Performing integration in (\ref{260}) in parts and taking
into account relation (\ref{280}) we arrive at,
\beqn
I_c(q) &\equiv& f_C^{(B)}(q)\,\Phi_C(q) = 
-\alpha Z_1Z_2\, f_C^{(B)}(q)\ 
\left[ \ln\left(\frac{q^2}{\lambda^2}\right) +
w\,\int\limits_{-1}^0 dt\,e^{-wt}\,
\ln\left|\frac{2+t}{t}\right|\right.\nonumber\\
&-& \left.w\,\int\limits_0^\infty dt\,e^{-wt}\,
\ln\left(\frac{2+t}{t}\right)\right] + 
O\left({\lambda^2\over q^2}\right)\ ,
\label{290}
\eeqn
 where
 \beq
w=\frac{a\,q^2}{2}\ .
\label{300}
\eeq
 The integrals contained in this expression can be represented in terms of
the integral exponential function,
\beqn
w\,\int\limits_{-1}^0 dt\,e^{-wt}\,
\ln\left|\frac{2+t}{t}\right| &=&
{\rm Ei}(w)-\gamma-\ln(2w)+
e^{2w}\,\Bigl({\rm E_1}(w)-{\rm E_1}(2w)\Bigr)\ ,
\nonumber\\
w\,\int\limits_0^\infty dt\,e^{-wt}\,
\ln\left(\frac{2+t}{t}\right) &=&
\gamma + \ln(2w) + e^{2w}\,{\rm E_1}(2w)\ .
\label{310}
\eeqn

Eventually we arrive at the following expression for the phase $\Phi_C(q)$,
\beq
\Phi_C(q) = \alpha Z_1Z_2\, 
\left\{2\,\ln(2w) - \ln\left(\frac{q^2}{\lambda^2}\right)
+2\,\gamma - {\rm Ei}(w) + e^{2w}\,
\Bigl[2\,{\rm E_1}(2w)-{\rm E_1}(w)\Bigr]\right\}\ .
\label{320}
\eeq

Apparently, the common phase factor for the terms in the elastic
amplitude (\ref{10}) is unobserved, only the difference
$\Delta\Phi=\Phi_C-\Phi_{NC}$, the so called Coulomb phase, matters,
 \beq
\Delta\Phi(q) =
\alpha Z_1Z_2\,\left\{\ln\left(\frac{a^2}{b^2}\right)
+{\rm Ei}(z)-{\rm Ei}(w) +
e^{2w}\,\Bigl[2\,{\rm E_1}(2w)-{\rm E_1}(w)\Bigr]\right\}\ .
\label{330}
\eeq
 Note that all the divergences and the photon mass $\lambda$ have
cancelled.

If the slopes of the Coulomb and hadronic formfactors coincide, $a=b$
($z=aq^2/2=w$), the expression (\ref{330}) takes an especially simple
form,
 \beq
\Delta\Phi(q) = 
\alpha Z_1Z_2\,e^{2w}\,\Bigl[2\,{\rm E_1}(2w)-{\rm E_1}(w)\Bigr]\ .
\label{340}
\eeq
 This is a good approximation for nuclei where the slopes of the
electromagnetic and hadronic formfactors are mainly determined by the nuclear
radius. For a proton target the electromagnetic slope
$a=4/\Lambda^2=5.6\,GeV^{-2}$. The slope of the hadronic amplitude $b$ is
approximately equal to $a$ at medium-high energies. However, it increases with
energy and at very high energies, in particular at the Tevatron, $b$
substantially exceeds $a$ and one should use the exact expression (\ref{330}).

\begin{figure}[tbh] 
\includegraphics{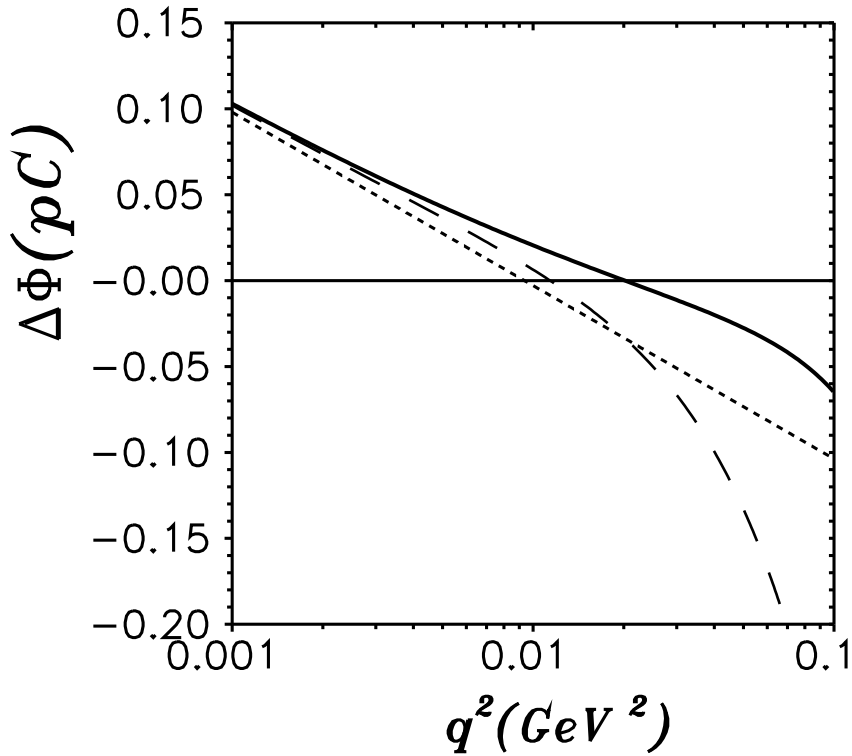} 
\begin{center} 
\vspace{10cm} 
\parbox{13cm}
{\caption[Delta]
{\sl The Coulomb phase as function of $q^2$. The dotted, dashed and solid 
curves corresponds to the formulas of Bethe \cite{bethe}, Cahn \cite{cahn}, 
and present calculations, respectively.} 
\label{phi}} 
\end{center} 
\end{figure}

To see the difference of our results from the previous calculations we
show $\Delta\Phi(q^2)$ by the solid curve in Fig.~\ref{phi}. We performed
calculations for proton-carbon elastic scattering using (\ref{340}) with
$a=30\,GeV^{-2}$.
 Dotted and dashed curves correspond to the formulas of Bethe
\cite{bethe} and Cahn \cite{cahn} (with $B=8/\Lambda^2=2a$) respectively.
At small $q^2$ dominated by the leading $\sim \ln(q^2)$  term
the results of Cahn and ours coincide and deviate from the Bethe's
due to the next-to-leading corrections $\sim q^2\ln(q^2)$.
At higher $q^2$ the next-to-next-to-leading corrections $\sim q^2$
become important and all three curves are quite different. 
This is not surprising since it is not legitimate to use the
results of \cite{bethe,cahn} at so large $q^2$.

Note that the eikonal approximation we use is subject to inelastic
corrections. In addition to the ordinary inelastic corrections to the
hadronic amplitude which we assume to be already included, excitation of
inelastic states $N^*$ between the electromagnetic and hadronic vertices
in Fig.~\ref{graphs}c can affect our results. Nevertheless, it turns out
that the correction to the Coulomb phase is quite small. Indeed, the
$N\to N^*$ amplitude contains an extra factor $q$ compared to the elastic
$N\to N$ one. At small $q\to 0$ this vertex squared cancels $1/q^2$ in
the photon propagator and this correction looks similar to the ordinary
inelastic one. We only have to replace factor $A^2/R_A^2$ in the hadronic
inelastic correction to $A\,Z\,\alpha/R_A^2$. Then, $\alpha Z$ is the
common factor for the Coulomb phase, and the relative correction becomes
$1/[(m_{N^*}^2-m_N^2)R_A^2]$. This is a small correction, about
$10^{-2}$, to the constant and $q^2R_A^2$ terms in the phase.

Concluding, we derived a new expression (\ref{330}) for the Coulomb phase
which is valid at any $q^2$ provided that the electromagnetic formfactor
and the hadronic amplitude have Gaussian dependences on $q$. If their
slopes are equal the Coulomb phase takes a very simple form (\ref{340}).

\medskip

\noindent {\bf Acknowledgment}: We are grateful to Misha Ryskin for
interesting discussion of the role of inelastic corrections and to Nigel
Buttimore who read the manuscript and made many improving suggestions.
This work has been supported by a grant from the Gesellschaft f\"ur
Schwerionenforschung Darmstadt (GSI), grant No.~HD~H\"UFT and by the
Federal Ministry BMBF grant No.~06 HD 954, by the grant
INTAS-97-OPEN-31696, and by the European Network: Hadronic Physics with
Electromagnetic Probes, Contract No.~FMRX-CT96-0008.

\end{document}